\documentclass[twocolumn,english]{article}
\usepackage[T1]{fontenc}
\usepackage{amsmath}
\usepackage{graphicx}

\makeatletter

\providecommand{\LyX}{L\kern-.1667em\lower.25em\hbox{Y}\kern-.125emX\@}

\usepackage{a4wide,graphicx}
\usepackage{setspace}
\usepackage{babel}
\makeatother
\begin{document}

\title{{\huge Constraints on Models for Proton-Proton Scattering from Multi-strange
Baryon Data }}

\author{{\small F.M. Liu$^{1,2}$, J. Aichelin$^{1}$, M. Bleicher$^{1}$,}
{\small H.J. Drescher$^{3}$, S. Ostapchenko$^{4,5}$, T. Pierog$^{1}$,
and K. Werner$^{1}$}\textbf{\emph{\small }}\\
 \textit{\small $^{1}$ SUBATECH, Université de Nantes -- IN2P3/CNRS
-- Ecole des Mines,  Nantes, France }\\
\textit{\small $^{2}$Institute of Particle Physics , Huazhong Normal
University, Wuhan, China}\\
 \textit{\small $^{3}$ Physics Department, New York University, New
York, USA} {\small }\textit{\small }\\
\textit{\small $^{4}$ Institut für Experimentelle Kernphysik, University
of Karlsruhe, 76021 Karlsruhe, Germany} {\small }\textit{\small }\\
\textit{\small $^{5}$ Moscow State University, Institute of Nuclear
Physics, Moscow, Russia  }}

\maketitle
\begin{abstract}
The recent data on pp collisions at 158 GeV provide severe constraints
on string models: These measurements allow for the first time to determine
how color strings are formed in ultrarelativistic proton-proton collisions. 
\end{abstract}

\section{Introduction}

\begin{figure}[htp]
\begin{center}\includegraphics[  width=0.30\paperwidth]{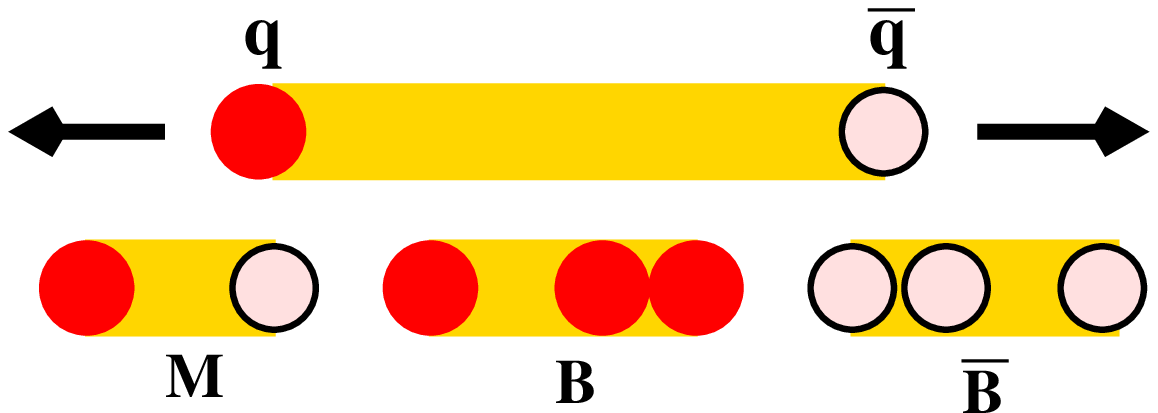}\end{center}

\caption{{\small \label{ee}$\mathrm{e}^{+}\mathrm{e}^{-}\rightarrow \gamma ^{*}\rightarrow \mathrm{q}\overline{\mathrm{q}}$.
The $\mathrm{q}-\overline{\mathrm{q}}$ string fragments into hadrons.}}
\end{figure}

Recently, the NA49 collaboration has published \cite{Kadija:sqm2001}
the rapidity spectra of p, $\Lambda $, $\Xi $ as well as the corresponding
antibaryons in pp interactions at 158 GeV. These measurements provide
new insight into the string formation process. In the string picture,
high energy proton-proton collisions create {}``excitations'' in
form of strings, being one dimensional objects which decay into hadrons
according to longitudinal phase space. This framework is well confirmed
in low energy electron-positron annihilation \cite{Werner:1993uh}
where the virtual photon decays into a quark-antiquark string which
breaks up into mesons($M$), baryons($B$) and antibaryons($\overline{B}$).
An example of a $\mathrm{q}-\overline{\mathrm{q}}$ string fragmenting
into hadrons is shown in Fig.\ref{ee}. Proton-proton collisions are
more complicated due to the fact that even at 158 GeV proton-proton
collisions are governed by soft physics, thus pQCD calculations can
not be applied. And the mechanism of string formation is not clear,
as will be discussed in the following.%

One may distinguish two classes of string models:

\begin{itemize}
\item Longitudinal excitation (LE) models: UrQMD \cite{Bleicher:1999xi},
HIJING \cite{Wang:1996yf}, PYTHIA \cite{Sjostrand:2000wi}, FRITIOF
\cite{Pi:1992ug}; 
\item Color exchange (CE) models: DPM \cite{Ranft:1988kc}, VENUS \cite{Werner:1993uh},
QGS \cite{Kaidalov:1982}.\vskip -0.2cm%
\begin{figure}[htp]
\begin{center}\includegraphics[  width=0.30\paperwidth]{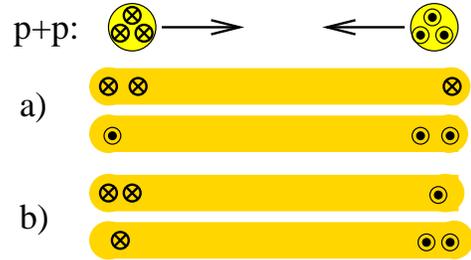}\end{center}

\caption{\label{pp} {\small Two string formation mechanisms for pp collisions
are presented: a) longitudinal excitation (LE). b) color exchange
(CE).}}
\end{figure}

\end{itemize}
In the LE case the two colliding protons excite each other via a large
transfer of momentum between projectile and target, Fig.\ref{pp}a.
In contrast, the CE picture considers a color exchange between the
incoming protons, leaving behind two octet states. Thus, a diquark
from the projectile and a quark from the target, and vice versa, form
color singlets. These are identified with strings, c.f. Fig.\ref{pp}b.
The color exchange is a soft process. The transfer of momentum is
negligible. The final result, two quark-diquark strings with valence
quarks being their ends, however, is quite similar. 

How are baryons and antibaryons produced? The easiest way to obtain
baryons is to break the strings via quark-antiquark pair production
close to the valence diquark. Since the ingoing proton was composed
of light quarks(qqq), the resulting baryon is of qqq or qqs type.
Thus nucleons, $\Lambda $s or $\Sigma $s are formed. Since these
baryons are produced at the string ends, they occur mainly close to
the projectile rapidity or target rapidity (leading baryons). 

Multi-strange baryons which consist of two or three strange quarks
are produced near the quark end or the middle of the strings, via
ss-${\overline{\mathrm{s}}}{\overline{\mathrm{s}}}$ production. Therefore
the distributions of multi-strange baryons are peaked around central
rapidity and the corresponding yields of multi-strange baryons and
their antiparticles should be comparable. A closer look reveals an
interesting phenomenon: Theoretically one finds the ratio of yields
\cite{Bleicher:2001nz}: \[
\overline{\Xi }^{+}/\Xi ^{-}=0.8\sim 1.2\, .\]
Experimentally, however, $\overline{\Xi }^{+}$s are less frequent
than expected. The ratio at midrapidity is \cite{Kadija:sqm2001}\[
\overline{\Xi }^{+}/\Xi ^{-}=0.44\pm 0.08.\]
 The situation for $\Omega $s is even more extreme: from string models
one gets \cite{Bleicher:2001nz}\[
\overline{\Omega }^{+}/\Omega ^{-}=1.6\sim 1.9\]
 at midrapidity. From extrapolating $\Lambda $ and $\Xi $ results
(and from preliminary NA49 data) we expect \cite{Kadija:sqm2001}\[
\overline{\Omega }^{+}/\Omega ^{-}=0.5\sim 0.8\, .\]
This is a generic situation; It is impossible to get the $\overline{\Omega }^{+}/\Omega ^{-}$
ratio smaller than unity from those two types of string models. As
addressed in \cite{Bleicher:2001nz}, this is due to the fact that
the strings have a light quark (but not a strange quark) at the end,
which disfavours multi-strange baryon production, and does not allow
for $\Omega $ production in the fragmentation region.

\section{Problems with the String Model Approach}

\begin{figure}[htp]
\begin{center}{\Large a)}$\; $\includegraphics[  scale=0.5]{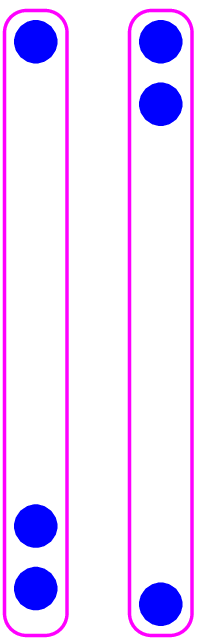}$\qquad \qquad ${\Large b)}$\: $\includegraphics[  scale=0.5]{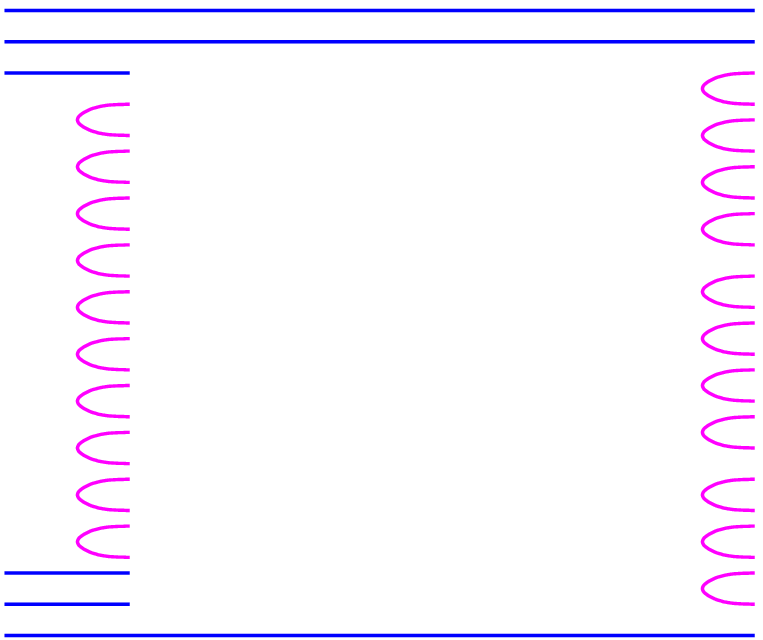}\end{center}

\caption{\label{chains}Two chains of quark lines (b) representing a pair
of strings (a)}
\end{figure}

\begin{figure}[htp]
\begin{center}{\Large a)}$\: $\includegraphics[  scale=0.5]{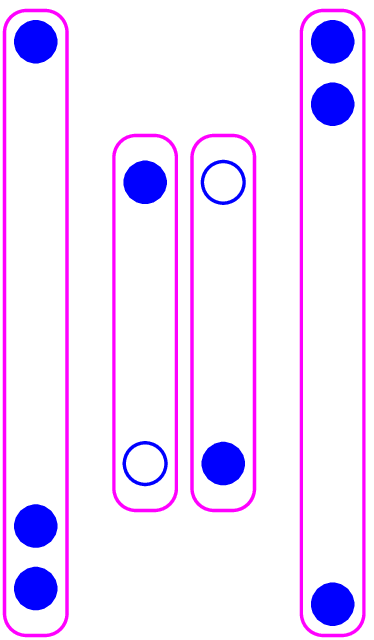}$\qquad \qquad ${\Large b)$\: $\includegraphics[  scale=0.5]{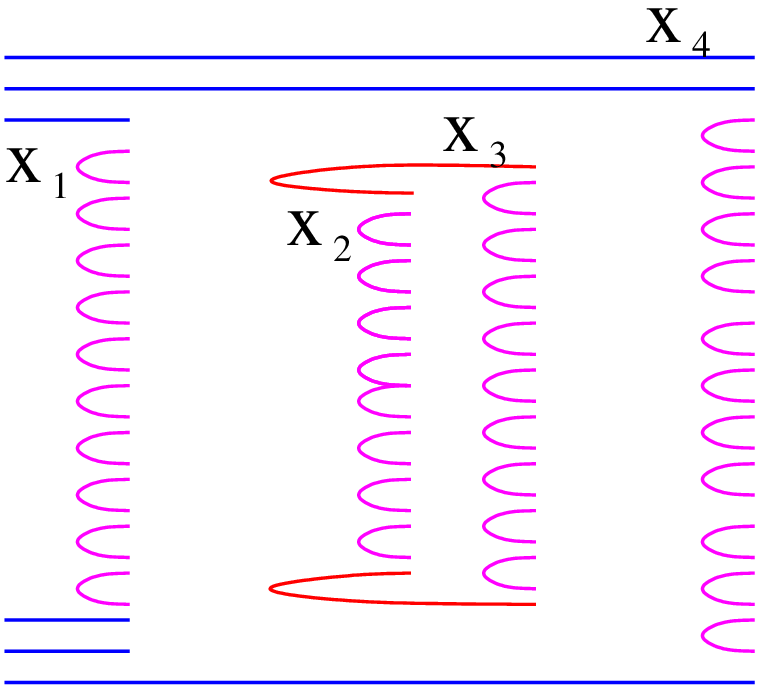}}\end{center}{\Large \par}

\caption{\label{cap:Two-pairs-of}Two pairs of strings (a) and the corresponding
chains (b)}
\end{figure}

So is there something fundamentally wrong with string models? To answer
this question let us consider somewhat more in detail how string models
are realized. One may present the particle production from strings
via chains of quark lines \cite{Ranft:1988kc} as shown in fig. \ref{chains}.
It turns out that the two string picture is not enough to explain
for example the large multiplicity fluctuations in proton-proton scattering
at collider energies: more strings are needed, one adds therefore
one or more pairs of quark-antiquark strings, as shown in fig. \ref{cap:Two-pairs-of}.%
 The variables $x_{i}$ refer to the longitudinal momentum fractions
given to the string ends. Energy-momentum conservation implies $\sum x_{i}=1.$

\begin{figure}[htp]
\begin{center}\includegraphics[  scale=0.6]{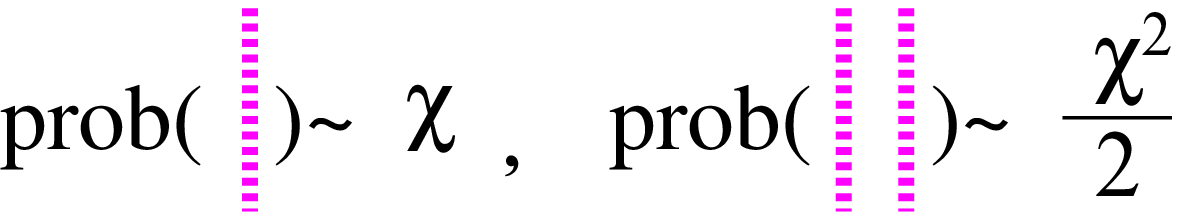}\end{center}

\caption{\label{gribov}Probabilities for configurations with one and two
elementary interactions (Pomerons), represented as dashed lines.}
\end{figure}

What is the probabilities for different string numbers? Here, Gribov-Regge
theory comes at help, which tells us that the probability for a configuration
with $n$ elementary interactions is given as\[
\mathrm{Prob}(n\, \mathrm{interactions})=\frac{\chi ^{n}}{n!}\exp (-\chi ),\]
where $\chi $ is a function of energy and impact parameter, see fig.
\ref{gribov}. %
Here, a dashed vertical line represents an elementary interaction
(referred to as Pomeron).

Now one identifies the elementary interactions (Pomerons) from Gribov-Regge
theory with the pairs of strings (chains) in the string model, and
one uses the above-mentioned probability for $n$ Pomerons to be the
probability for configurations with $n$ string pairs. Unfortunately
this is not at all consistent, for two reasons:

\begin{enumerate}
\item Whereas in the string picture the first and the subsequent pairs are
of different nature, in the Gribov-Regge model all the Pomerons are
identical.
\item Whereas in the string (chain) model the energy is properly shared
among the strings, in the Grivov-Regge approach does not consider
energy sharing at all (the $\chi $ is a function of the total energy
only)
\end{enumerate}
These problems have to be solved in order to make reliable predictions.

\section{Basic Ideas of Parton-Based Gribov-Regge Theory}

\begin{figure}[htp]
\begin{center}\includegraphics[  scale=0.5]{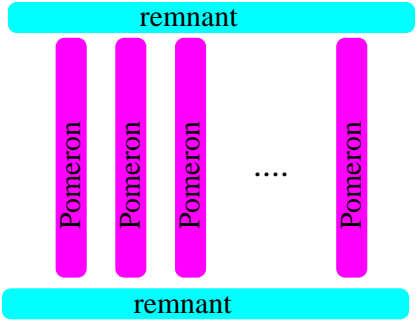}\end{center}

\caption{\label{mult} {\small Multiple elementary interactions (Pomerons)
in NEXUS.}}
\end{figure}

\begin{figure}[htp]
\begin{center}\includegraphics[  scale=0.6]{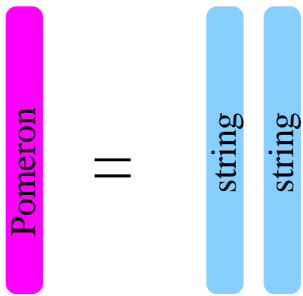}\end{center}

\caption{\label{pomstri} {\small Multiple elementary interaction (Pomerons)
in NEXUS (a). An elementary interaction is identified with a pair
of strings.}}
\end{figure}

We consider a new approach called Parton-Based Gribov-Regge theory
to solve the above-mentioned problems. Here we still use the language
of Pomerons as in Gribov-Regge theory to calculate probabilities of
collision configurations and the language of strings to treat particle
production. Multiple interactions happen in parallel. An elementary
interaction is referred to as Pomeron. The spectators of each proton
form a remnant, see Fig. \ref{mult}. A Pomeron is finally identified
with two strings, see Fig.\ref{pomstri}. But we treat both aspects
in a consistent fashion: \emph{In both cases energy sharing is considered
in a rigorous way}\cite{Drescher:2000ha}\emph{, and in both cases
all Pomerons are identical.} This is the new feature of our approach,
and the second point is exactly the solution to multistrange baryon
problem mentioned above. In the following we discuss how to realize
the collision configuration and particle production, respectively.%

\section{Collision Configuration}

Gribov-Regge theory is applied in {\small NE}{\large X}{\small US
to calculate collision configurations.} This calculation is performed
under the condition that \emph{energy sharing is considered rigorous}ly.
Once a collision configuration is determined, not only the number
of Pomerons but also the energy sharing among Pomerons is fixed. This
is different from other string models. \emph{}In the following we
discuss how to realize it in a qualitative fashion.

\subsection{Reminder: some Elementary Quantum Mechanics}

\begin{figure}[htp]
\begin{center}\includegraphics[  height=0.10\paperwidth]{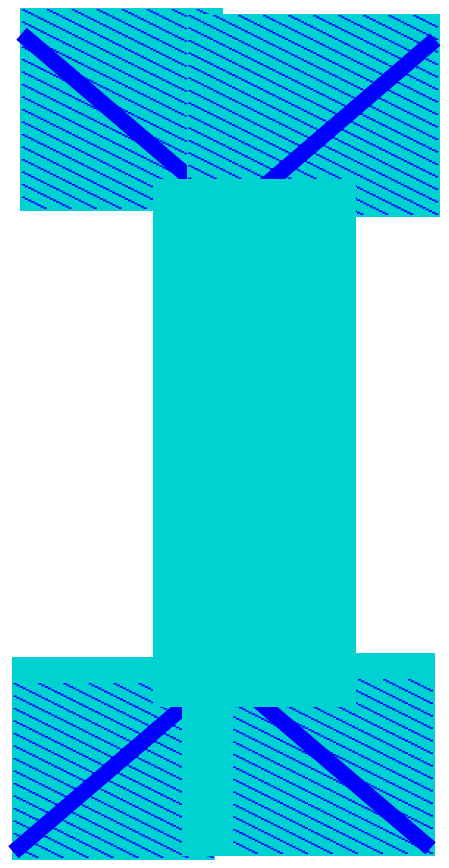}$\qquad $\includegraphics[  height=0.10\paperwidth]{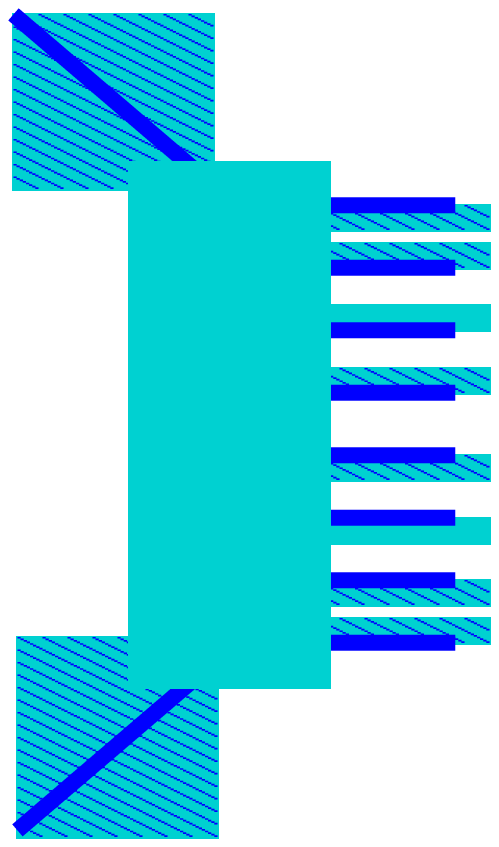}\end{center}

\caption{An elastic scattering amplitude $T_{2\rightarrow 2}$ (left) and
an inelastic amplitude $T_{2\rightarrow X}$ (right).\label{t0}}
\end{figure}

\begin{figure}[htp]
\begin{center}\includegraphics[  height=0.10\paperwidth]{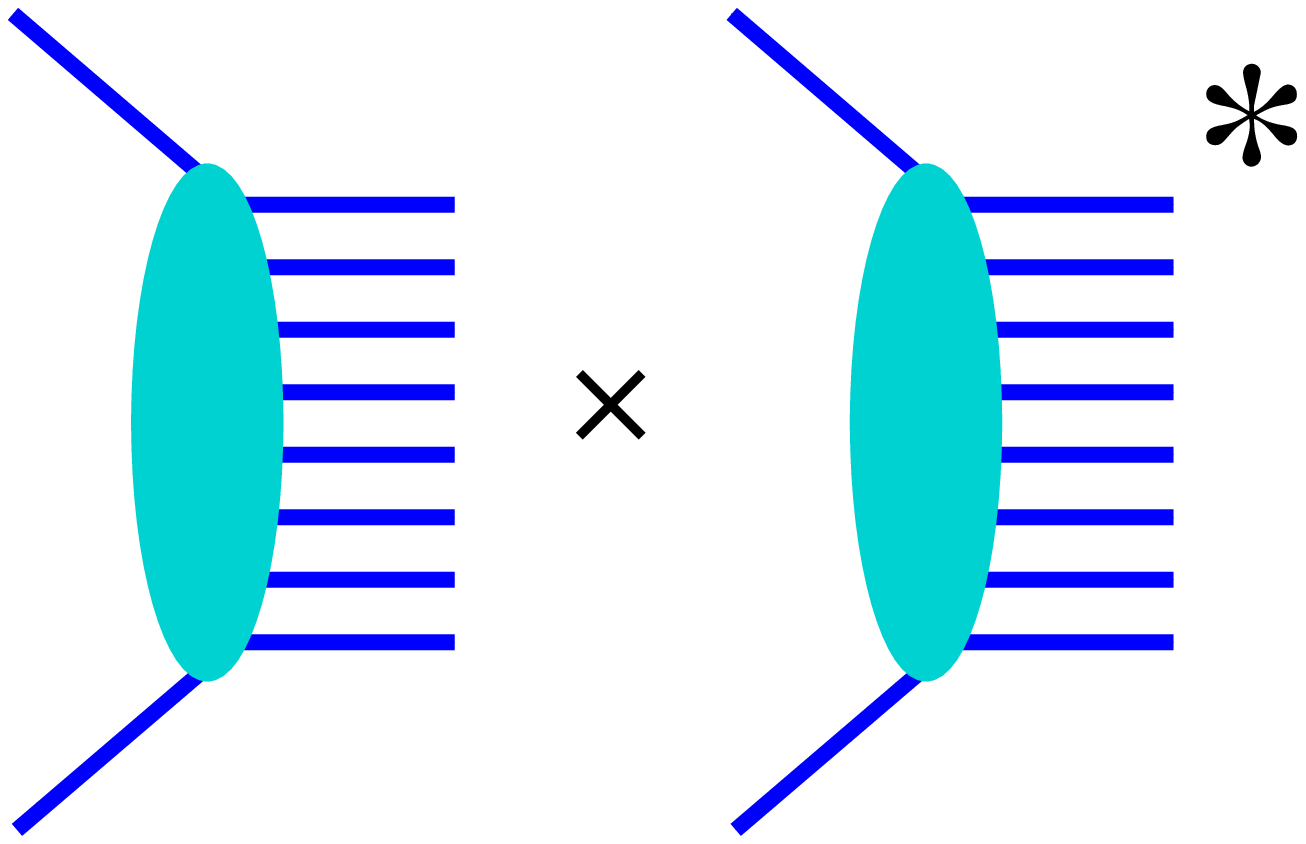}\includegraphics[  height=0.10\paperwidth]{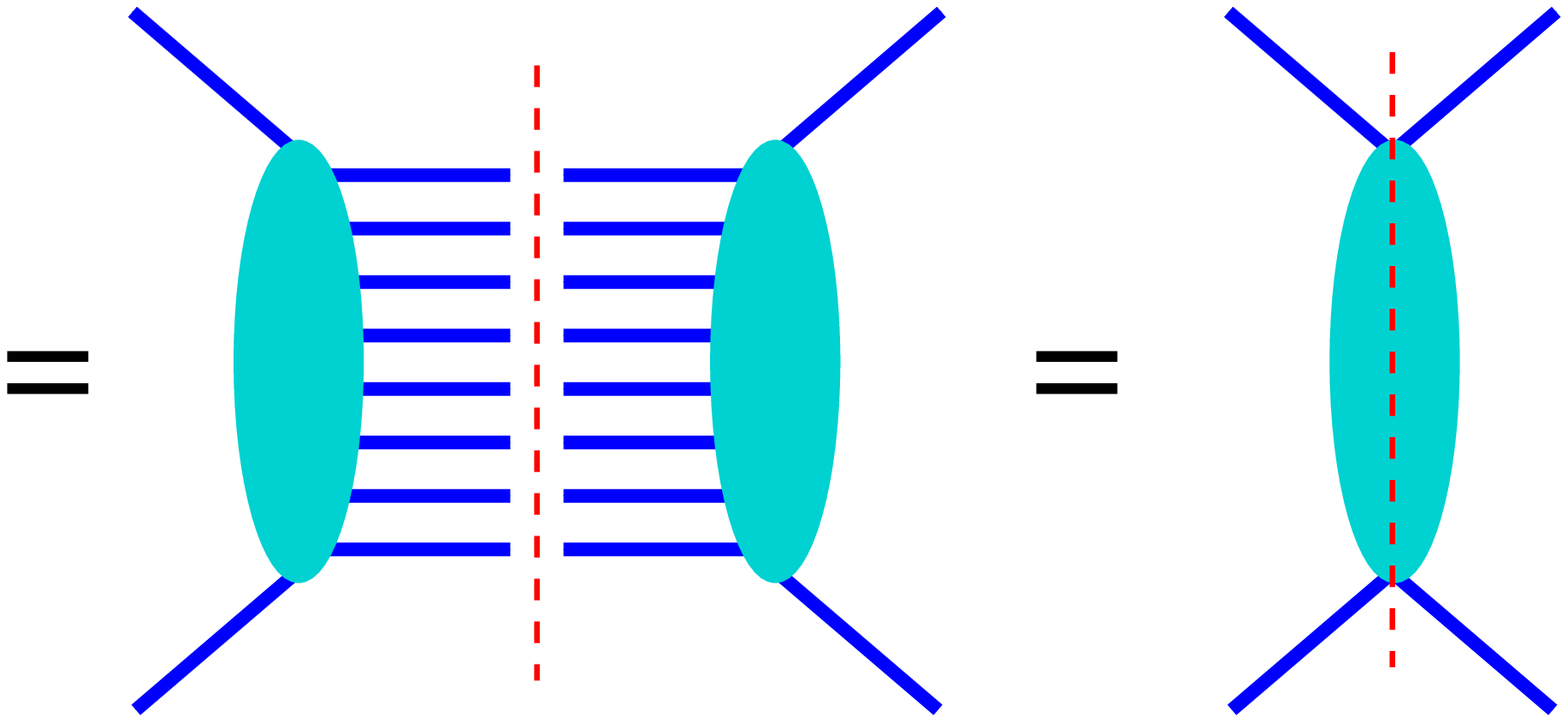}\end{center}

\caption{The expression $\sum _{X}(T_{2\rightarrow X})$$(T_{2\rightarrow X})^{*}$which
may be represented as a {}``cut diagram''.\label{t2}}
\end{figure}

\begin{figure}[htp]
\begin{center}\includegraphics[  scale=0.28]{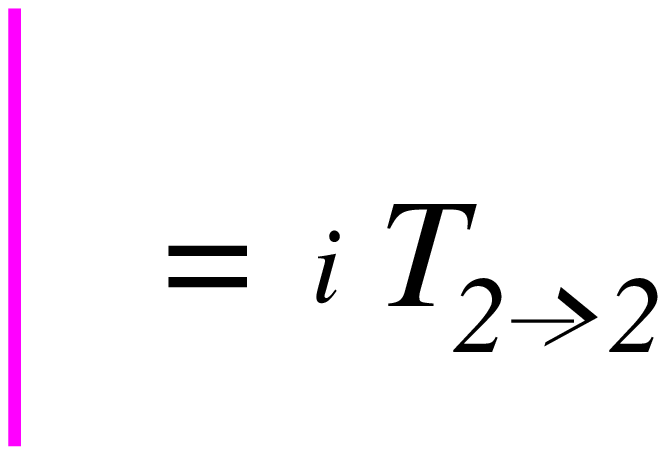},~~\includegraphics[  scale=0.28]{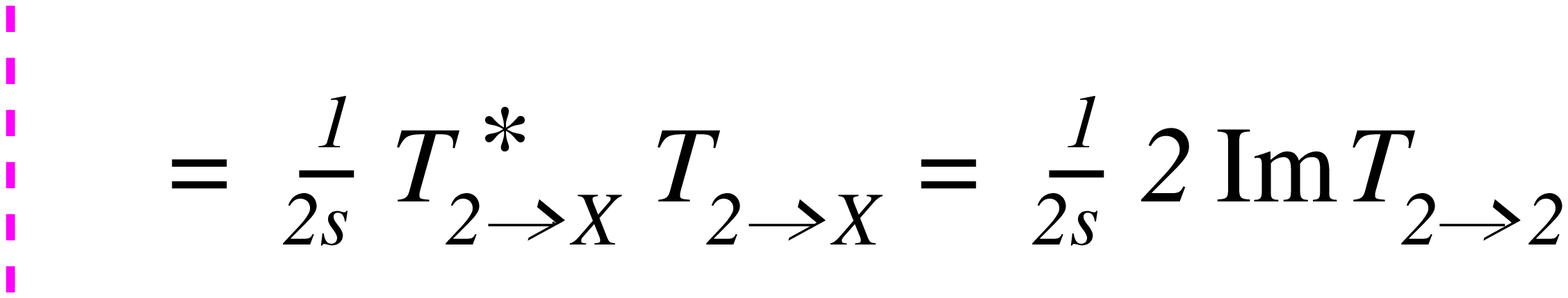}\end{center}

\caption{Conventions.\label{line}}
\end{figure}

Let us introduce some conventions. We denote elastic two body scattering
amplitudes as $T_{2\rightarrow 2}$ and inelastic amplitudes corresponding
to the production of some final state $X$ as $T_{2\rightarrow X}$
(see fig.\ref{t0} ). %
As a direct consequence of unitarity on has $2\, \mathrm{Im}T_{2\rightarrow 2}=\sum _{x}$$(T_{2\rightarrow X})$$(T_{2\rightarrow X})^{*}$.
The right hand side of this equation may be literally presented as
a {}``cut diagram'', where the diagram on one side of the cut is
$(T_{2\rightarrow X})$ and on the other side $(T_{2\rightarrow X})^{*}$,
as shown in fig.\ref{t2} . %
So the term {}``cut diagram'' means nothing but the square of an
inelastic amplitude, summed over all final states, which is equal
to twice the imaginary part of the elastic amplitude. Based on these
considerations, we introduce simple graphical symbols, which will
be very convenient when discussing multiple scattering, shown in fig.
\ref{line}: a vertical solid line represents an elastic amplitude
(multiplied by $i$, for convenience), and a vertical dashed line
represents the mathematical expression related to the above-mentioned
cut diagram (divided by $2s$, for convenience). %

\subsection{Elementary Interactions}

\begin{figure}[htp]
\begin{center}\includegraphics[  height=0.2\paperwidth]{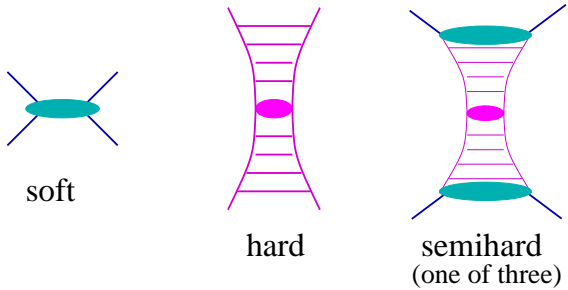}\end{center}

\caption{The elastic amplitude $T_{2\rightarrow 2}$.\label{t22}}
\end{figure}

Elementary nucleon-nucleon scattering can be considered as a straightforward
generalization of photon-nucleon scattering: one has a hard parton-parton
scattering in the middle, and parton evolutions in both directions
toward the nucleons. We have a hard contribution $T_{\mathrm{hard}}$
when the the first partons on both sides are valence quarks, a semi-hard
contribution $T_{\mathrm{semi}}$ when at least on one side there
is a sea quark (being emitted from a soft Pomeron), with a perturbative
process happening in the middle of Pomeron, and finally we have a
soft contribution when no perturbative process happens at all (see
fig. \ref{t22}). The total elementary elastic amplitude $T_{2\rightarrow 2}$
is the sum of all contributions.%

The cut-off of virtuality between perturbative and non-perturbative
processes, $Q_{0}^{2}$, is independent of the beam energy. The perturbative
part of Pomeron is strictly based on the standard DGLAP evolution
with ordered parton virtualities in the ladder diagram and the nonperturbative
part is based on Reggeon parameterization. The semihard Pomeron is
a convolution of hard and soft. The total elementary elastic amplitude
$T_{2\rightarrow 2}$ is the sum of all these terms. Thus we have
a smooth transition from soft to hard physics: at low energies the
soft contribution dominates, at high energies the hard and semi-hard
ones, at intermediate energies (that is where experiments are performed
presently) all contributions are important. 

The multiple scattering theory will be based on these elementary interactions,
the corresponding elastic amplitude $T_{2\rightarrow 2}$ and the
corresponding cut diagram, both being represented graphically by a
solid and a dashed vertical line. We also refer to the solid line
as Pomeron, to the dashed line as cut Pomeron.

\subsection{Multiple Scattering }

\begin{figure}[htp]
\begin{center}\includegraphics[  height=0.1\paperwidth]{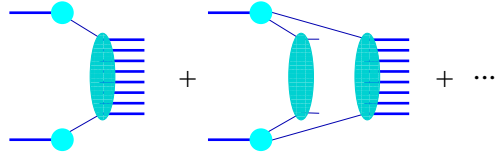}\end{center}

\caption{Inelastic scattering in pp. \label{t7}}
\end{figure}

\begin{figure}[htp]
\begin{center}(a)\includegraphics[  height=0.1\paperwidth]{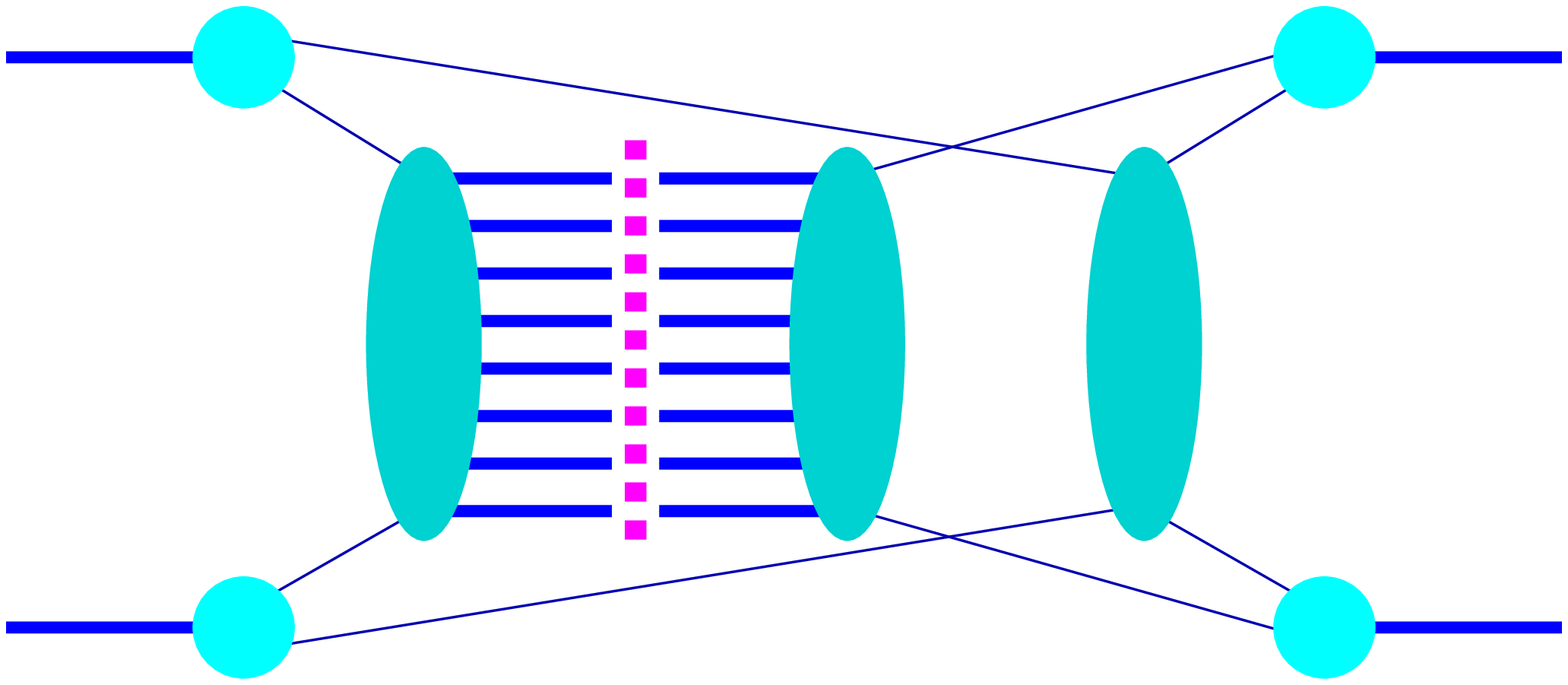}$\, \, $(b)\includegraphics[  height=0.1\paperwidth]{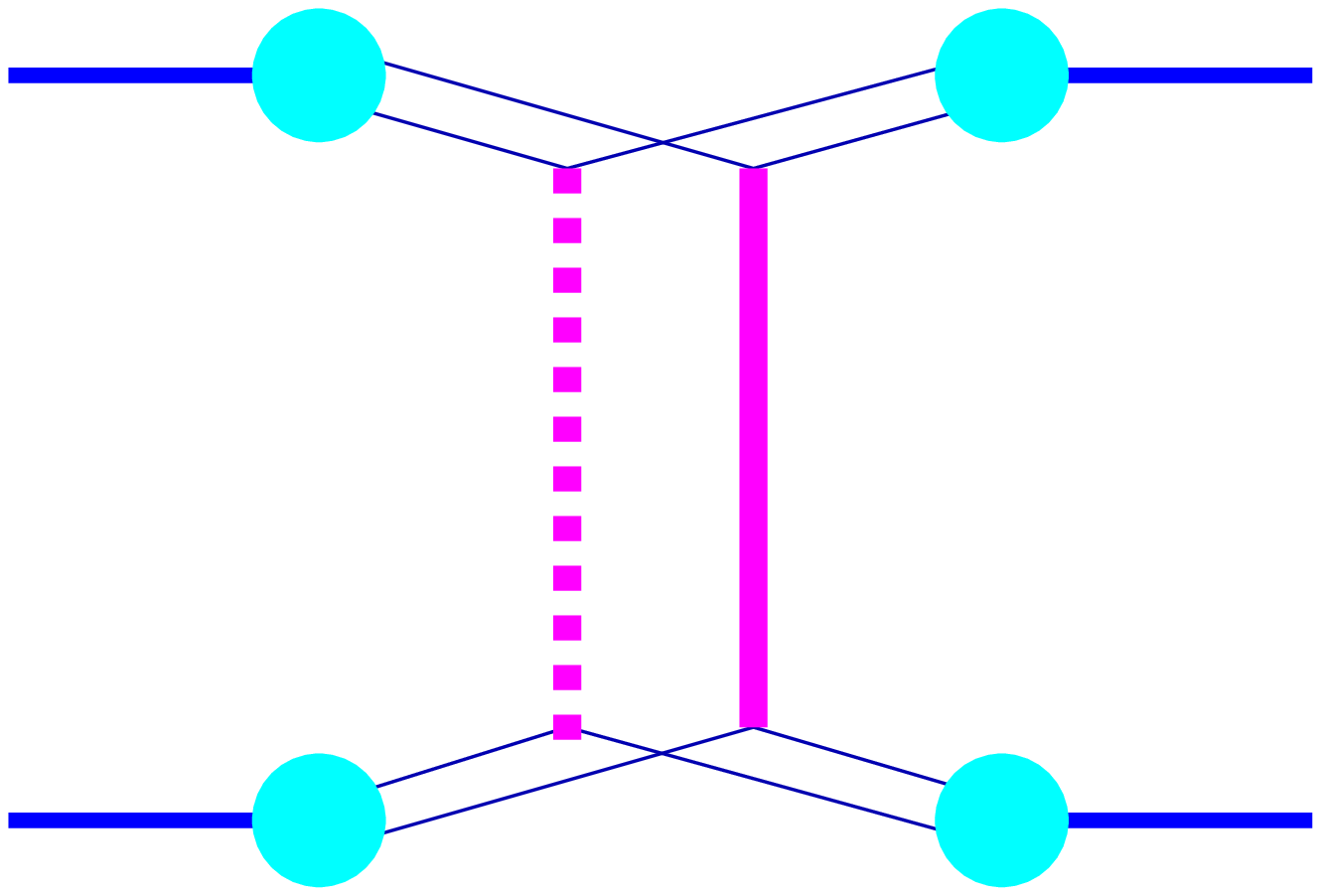}\end{center}

\caption{Inelastic scattering in pp.  a) An interference term of cross section,
b) Represented with our simplified notations.\label{t7b}}
\end{figure}

We first consider inelastic proton-proton scattering, see fig. \ref{t7}.
We imagine an arbitrary number of elementary interactions to happen
in parallel, where an interaction may be elastic or inelastic. The
inelastic amplitude is the sum of all such contributions with at least
one inelastic elementary interaction involved. %
To calculate cross sections, we need to square the amplitude, which
leads to many interference terms, as the one shown in fig. \ref{t7b}(a),
which represents interference between the first and the second diagram
of fig. \ref{t7}. Using the above notations, we may represent the
left part of the diagram as a cut diagram, conveniently plotted as
a dashed line, see fig. \ref{t7b}(b). The amplitude squared is now
the sum over many such terms represented by solid and dashed lines.%

\begin{figure}[htp]
\begin{center}\includegraphics[  height=1.5cm]{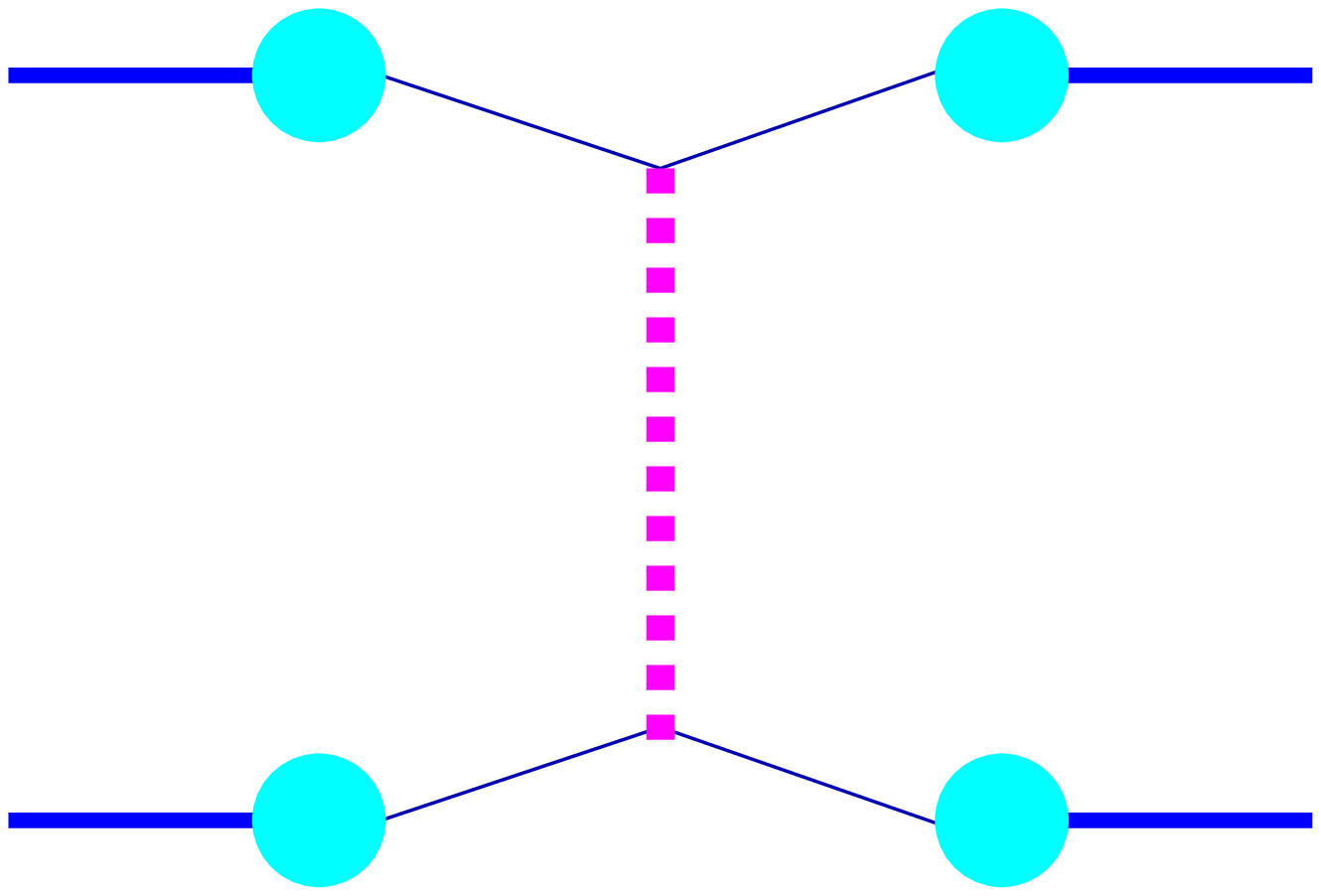}~$\, $\includegraphics[  height=1.5cm]{t7}$\, $~\includegraphics[  height=1.5cm]{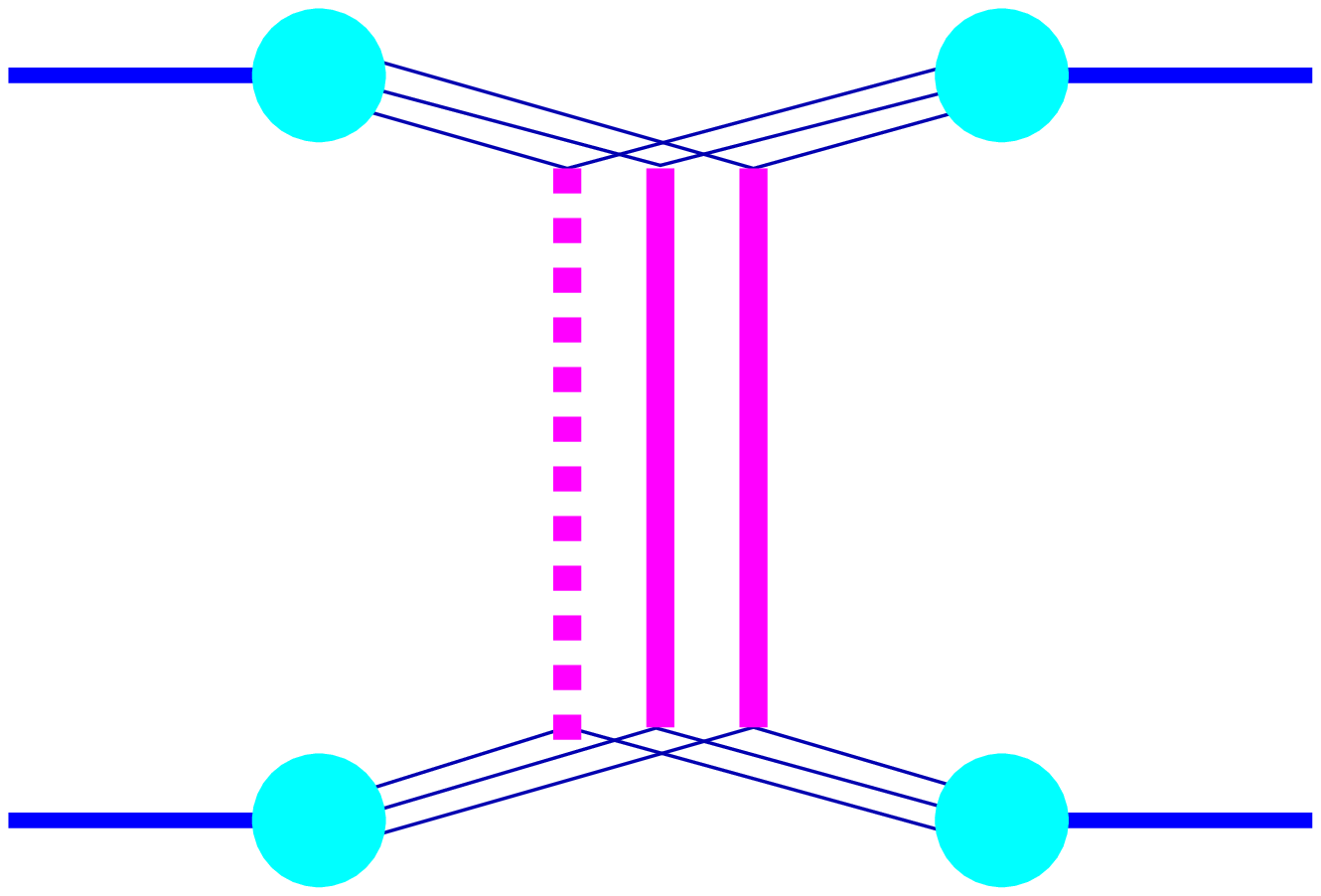}\end{center}

\caption{Class of terms corresponding to one inelastic interaction.\label{t7c}}
\end{figure}

\begin{figure}[htp]
\begin{center}\includegraphics[  height=1.5cm]{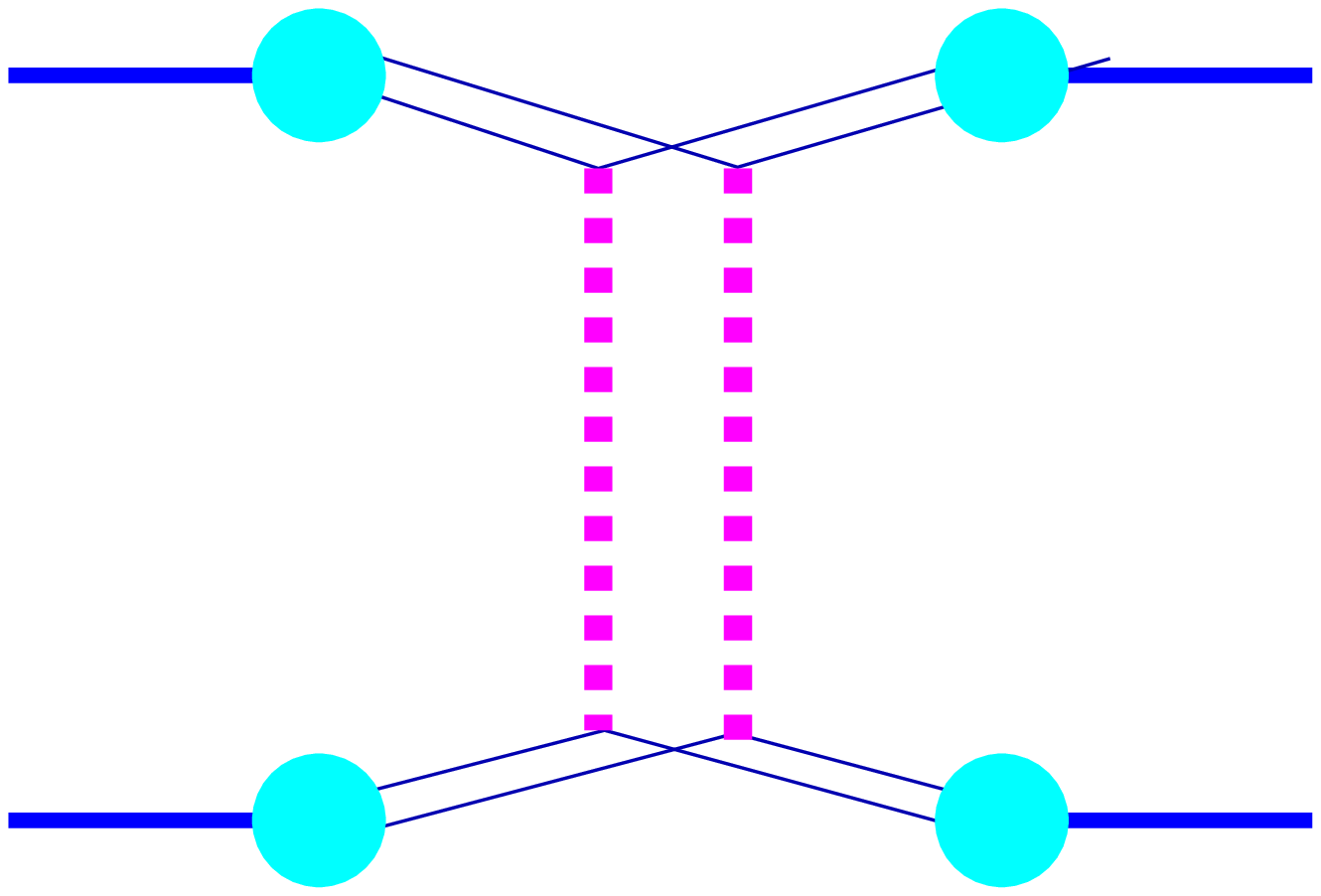}~$\, $\includegraphics[  height=1.5cm]{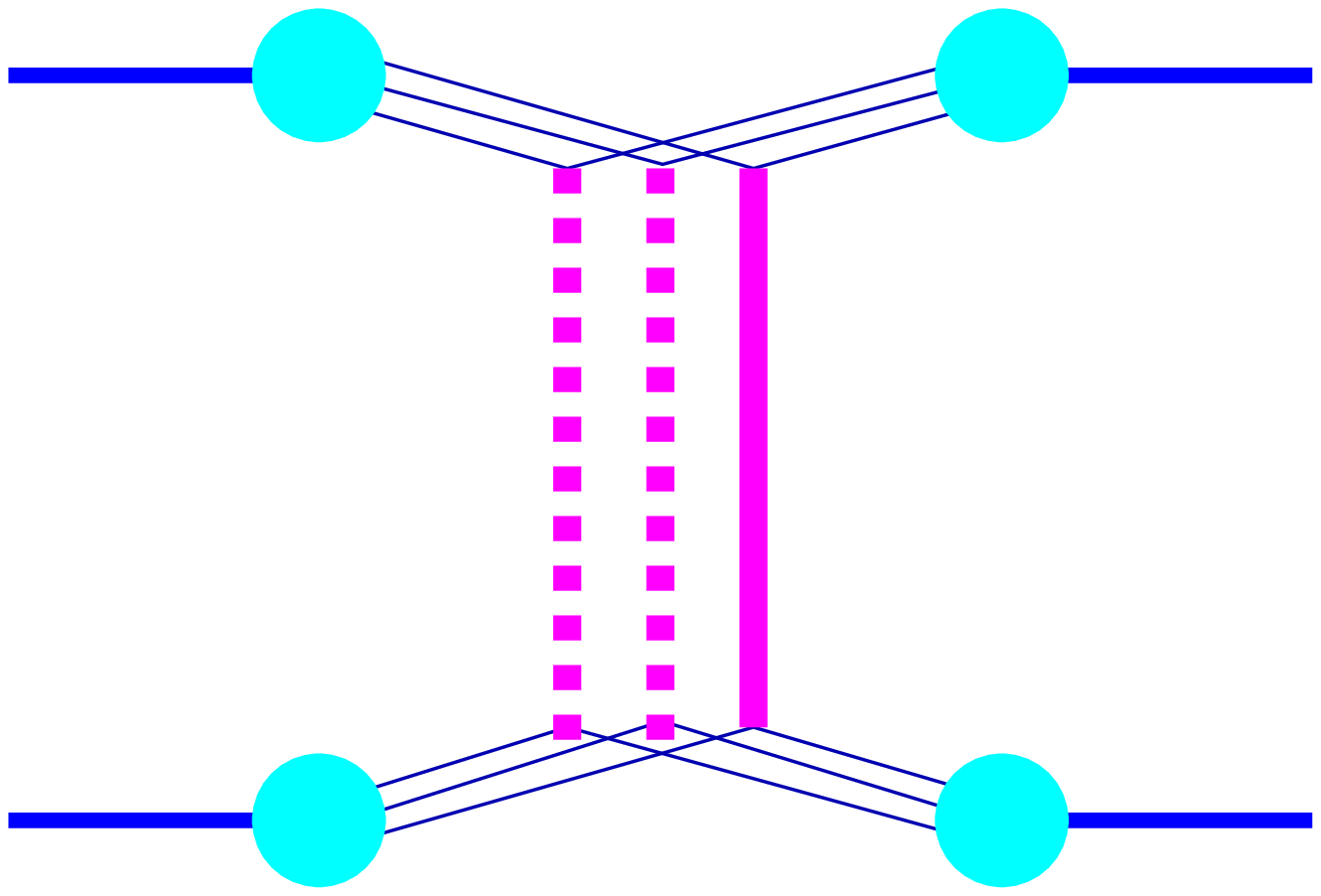}~$\, $\includegraphics[  height=1.5cm]{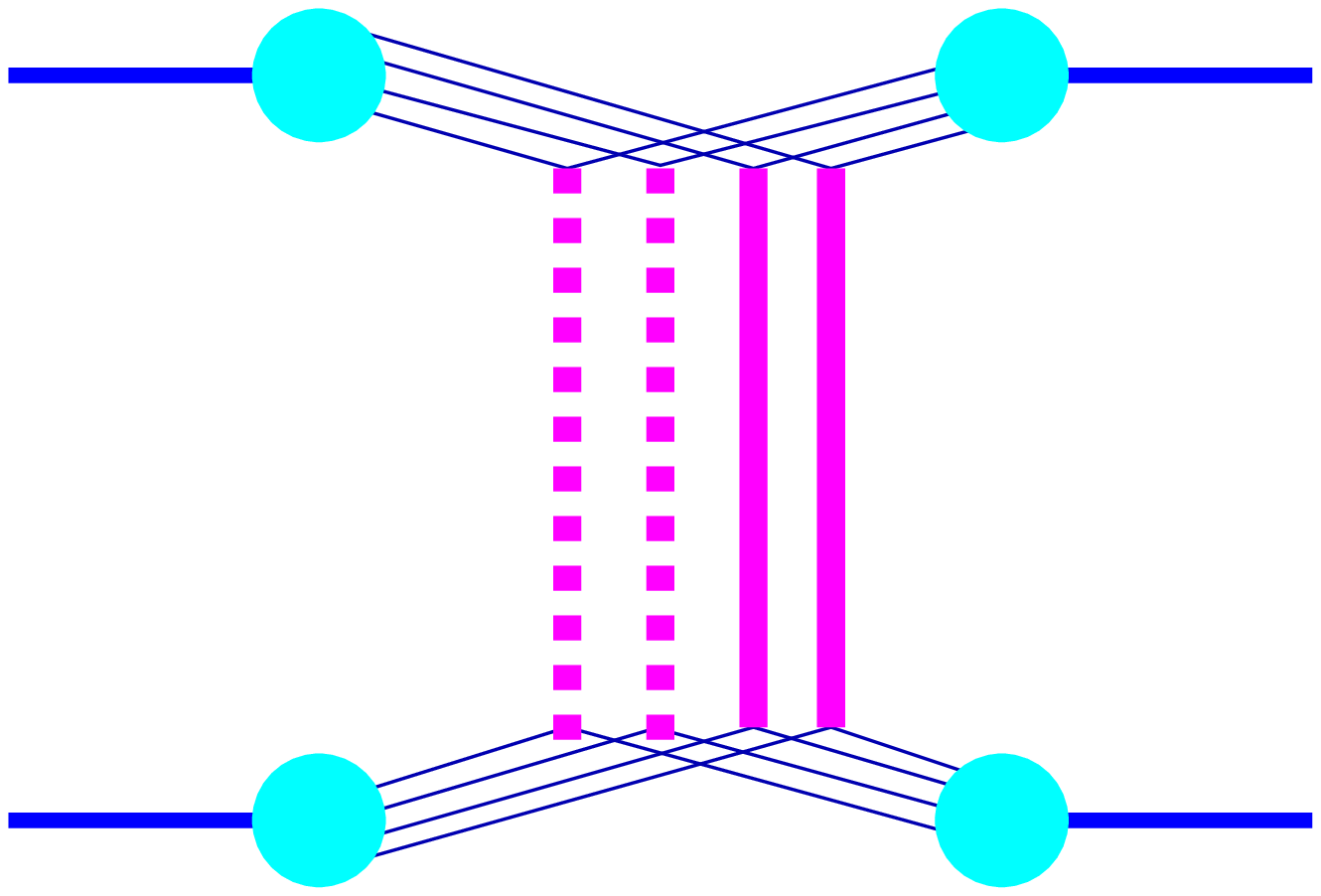}\end{center}

\caption{Class of terms corresponding to two inelastic interactions.\label{t8c}}
\end{figure}

When squaring an amplitude being a sum of many terms, not all of the
terms interfere -- only those which correspond to the same final state.
For example, a single inelastic interaction does not interfere with
a double inelastic interaction, whereas all the contributions with
exactly one inelastic interaction interfere. So considering a squared
amplitude, one may group terms together representing the same final
state. In our pictorial language, this means that all diagrams with
one dashed line, representing the same final state, may be considered
to form a class, characterized by $m=1$ -- one dashed line ( one
cut Pomeron) -- and the light cone momenta $x^{+}$ and $x^{-}$ attached
to the dashed line (defining energy and momentum of the Pomeron).
In fig. \ref{t7c}, we show several diagrams belonging to this class,
in fig. \ref{t8c}, we show the diagrams belonging to the class of
two inelastic interactions, characterized by $m=2$ and four light-cone
momenta $x_{1}^{+}$, $x_{1}^{-}$, $x_{2}^{+}$, $x_{2}^{-}$.%
Generalizing these considerations, we may group all contributions
with $m$ inelastic interactions ($m$ dashed lines = $m$ cut Pomerons)
into a class characterized by the variable\[
K=\{m,x_{1}^{+},x_{1}^{-},\cdots ,x_{m}^{+},x_{m}^{-}\}.\]
 We then sum all the terms in a class $K$,\[
\Omega (K)=\sum \{\mathrm{all}\, \mathrm{terms}\, \mathrm{in}\, \mathrm{class}\, K\}.\]
The inelastic cross section is then simply a sum over classes,\[
\sigma _{\mathrm{inel}}(s)=\sum _{K\neq 0}\int d^{2}b\, \Omega (K).\]
$\Omega $ depends implicitly on the energy squared $s$ and the impact
parameter $b$. The individual terms $\int d^{2}b\, \Omega (K)$,
represent partial cross sections, since they represent distinct final
states. They are referred to as topological cross sections. One can
prove\[
\sum _{K}\Omega (K)=1,\]
which is a very important result justifying our interpretation of
$\Omega (K)$ to be a probability distribution for the configurations
$K$. This provides also the basis for applying Monte Carlo techniques.

The function $\Omega $ is the basis of all applications of this formalism.
It is the foundation not only for calculating (topological) cross
sections, but also for particle production, thus providing a consistent
formalism for all aspects of a nuclear collision.

\section{Particle Production }

Here we discuss how to produce hadrons with any given collision configuration
determined according to the above section. As we discussed above,
each cut Pomeron is identified as a pair of strings. Similar to other
string models, a Lorentz invariant string fragmentation procedure
provides a transformation from strings to final-state hadrons in {\small NE}{\large X}{\small US}.
However the string formation mechanism in {\small NE}{\large X}{\small US}
is very different from others because Pomerons are treated identically
for calculating particle production.

\subsection{Hard Pomerons }

As discussed above, each elementary interaction has hard, soft and
semihard contribution. So each Pomeron has certain probabilities to
be of type soft, hard and semihard. Once a collision configuration
is determined according to the above section, the number of each type
of Pomerons is fixed. To give a proper description of deep inelastic
scattering data, hard and some of semihard Pomerons are connected
to the valance quarks of the hadron. In order to conserve the initial
hadron baryon content and to keep the simple factorized structure
with the leading logarithmic approximation of quantum Chromodynamics,
we associate a {}``quasi-spectator'' antiquark (of the same flavour)
to each valence quark interaction. So a hard Pomeron is a two-layer
ladder diagram where one of them is hard with ordered virtualities
and the other is soft. The hard ladder diagram gives a kinky string
(each string segment has a constant velocity and finally is identified
with those emitted perturbative partons). The fraction of hard Pomerons
in total Pomerons is very subdominant in average proton-proton collisions.
At SPS energies hard and semihard Pomerons do not contribute at all,
and even at collider energies the Pomerons connected to valence quarks
are rare. Therefore, even though all types of Pomerons are included
in the Monte Carlo, their contribution can be ignored in the following
discussion.

\subsection{Soft Pomerons}

How to form strings from soft Pomerons? No matter whether single-Pomeron
exchange or multiple-Pomeron exchange happens in a proton-proton scattering,
all Pomerons are treated identically. Because of this, it is a natural
idea to take quarks and antiquarks from the sea as string ends{\small ,}
because an arbitrary number of Pomerons may be involved, and valence
quarks are not always available to be string ends due to their limited
amount. {\small }This point is different from the above-mentioned
string models, where all the string ends are valence quarks. Letting
all the valence quarks stay in remnants, thus, string ends from cut
Pomerons have complete flavour symmetry and produce particles and
antiparticles in equal amounts. In the old version of {\small NE}{\large X}{\small US}
used in \cite{Bleicher:2001nz}, a sea quark and an antiquark of the
same flavour have been taken from the sea as string ends to keep flavour
conservation. In the new version, {\small NE}{\large X}{\small US
3}.0, the flavours of the string ends are independent. In order to
compensate the flavour, whenever a quark or an antiquark is taken
from the sea as a string end, a corresponding antiparticle from the
sea is put in the remnant nearby.

\subsection{Remnant }

Remnants are a new object, compared to other string models. The partonic
content of a remnant is very clear: three valence quarks and the compensated
partons. From this partonic structure one can estimate that its mass
is of the order of proton mass. The 4-momentum of a remnant is that
of an initial proton minus those being taken away by participant partons.
So the 4-momenta of projectile remnant and target remnant are fixed
once the collision configuration is given. The remnant mass can be
calculated according to its 4-momentum and the on-shell condition,
however this value of mass is not reliable. During the configuration
calculation one distributes energy of order 100 GeV to Pomerons and
remnants where a proton mass is negligible. We justify the remnant
mass with a distribution $P(m^{2})\propto (m^{2})^{-\alpha }$, $m^{2}\in (m_{\mathrm{min}}^{2},\, x^{+}s)$,
where $s$ is the squared energy at center mass system, $m_{\mathrm{min}}$
is the minimum mass of hadrons to be made from the remnant's quarks
and antiquarks, and $x^{+}$ is the light-cone momentum fraction of
the remnant which is determined in the collision configuration. Through
fitting the data we determine the parameter $\alpha =1.5$. Remnants
decay into hadrons according to n-body phase space\cite{droplet}.

\subsection{Leading Order Discussion}

\begin{figure}[htp]
\begin{center}\includegraphics[  width=0.34\paperwidth]{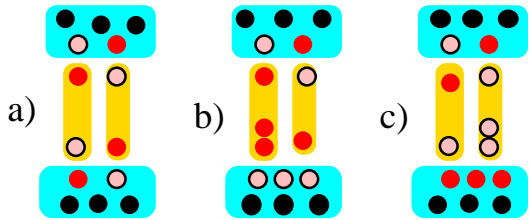}\end{center}

\caption{\label{nexus2} {\small a) The most simple and frequent collision
configuration has two remnants and only one cut Pomeron represented
by two $\mathrm{q}-\overline{\mathrm{q}}$ strings. b) One of the
$\overline{\mathrm{q}}$ string-ends can be replaced by a $\mathrm{qq}$
string-end. c) With the same probability, one of the $\mathrm{q}$
string-ends can be replaced by a $\overline{\mathrm{q}}\overline{\mathrm{q}}$
string-end. }}
\end{figure}

The configuration of leading order has only one cut Pomeron. The most
simple and most frequent collision configuration has two remnants
and only one cut Pomeron represented by two $\mathrm{q}-\overline{\mathrm{q}}$
strings as in Fig.\ref{nexus2}a. Besides the three valence quarks,
each remnant has additionally a quarks and an antiquark to compensate
the flavour. %

The leg of a cut Pomeron may be of qqq type with small probability
$P_{\mathrm{qq}}$, which means the corresponding string ends are
a diquark and a quark. In this way we get quark-diquark (q-qq) strings
from cut Pomerons. The qqq Pomeron leg has to be compensated by the
three corresponding antiquarks in the remnant, as in Fig.\ref{nexus2}b.
The (3q3$\overline{\mathrm{q}}$) remnant may decay into three mesons
(3M) or a baryon and an anti-baryon (B+$\overline{\mathrm{B}}$).
Since the 3M mode is favored by phase space, we neglect B+$\overline{\mathrm{B}}$
production here. 

For symmetry reasons, the leg of a cut Pomeron is of $\overline{\mathrm{q}}\overline{\mathrm{q}}\overline{\mathrm{q}}$
type with the same probability $P_{\mathrm{qq}}$. This yields a $\overline{\mathrm{q}}-\overline{\mathrm{q}}\overline{\mathrm{q}}$
string and a (6q) remnant, as shown in Fig.\ref{nexus2}c. The (6q)
remnant decays into two baryons. Since q-qq strings and $\overline{\mathrm{q}}-\overline{\mathrm{q}}\overline{\mathrm{q}}$
strings have the same probability to appear from cut Pomerons, baryons
and antibaryons are produced equally. However, from remnant decay,
baryon production is favored due to the initial valence quarks.

\section{Results}

Here, we will concentrate on baryon-antibaryon production, because
there we obtain strikingly different results compared to other models.
However, we carefully checked as well mesons -- essentially pion and
kaon rapidity and transverse momentum spectra, where the results are
quite close to what we obtained earlier with NEXUS 2 or VENUS.

\begin{figure}[htp]
\begin{center}\includegraphics[  width=0.80\paperwidth]{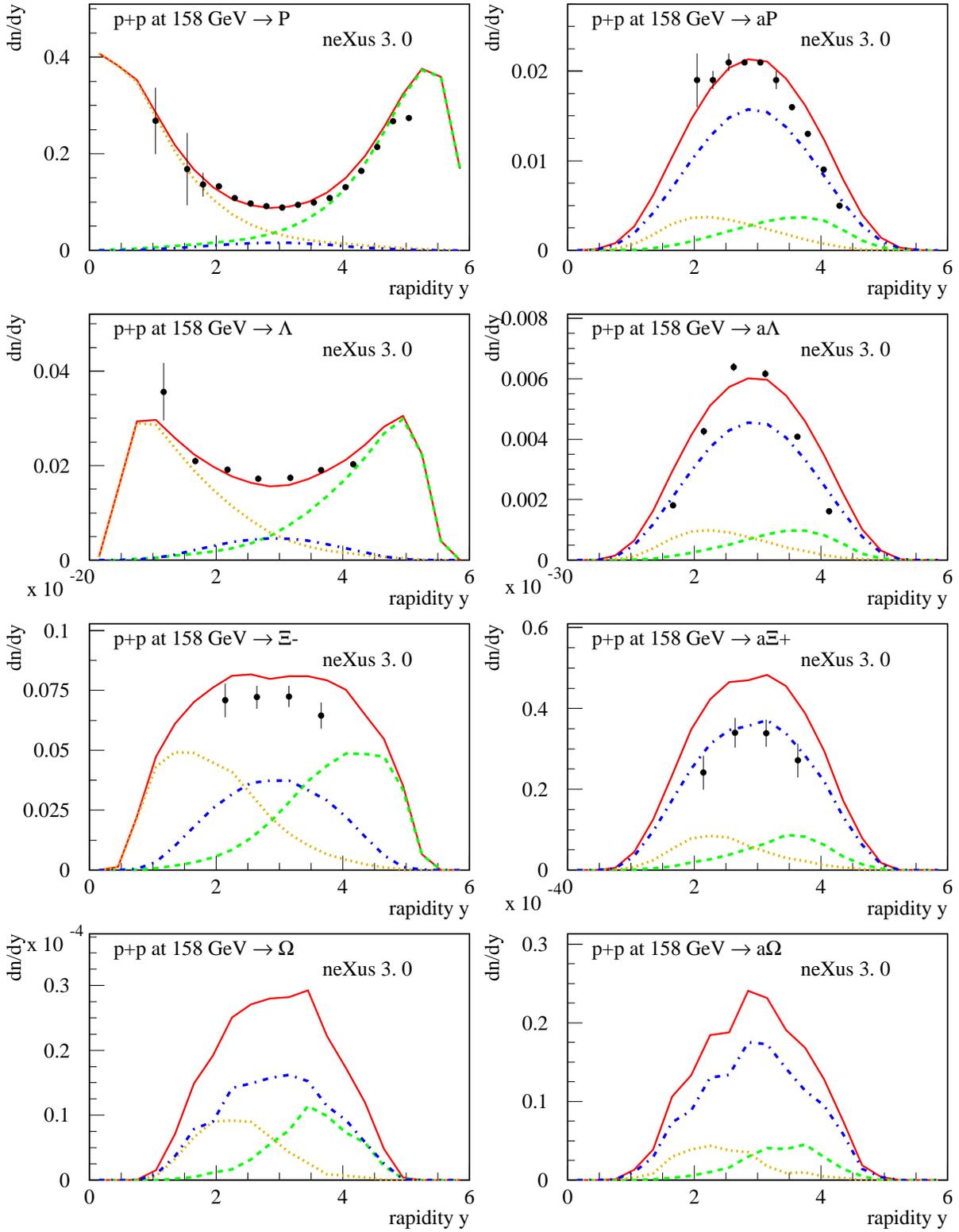}\end{center}

\caption{{\small \label{nexus3}Rapidity spectra of baryons and antibaryons
calculated from NEXUS 3.0 (projectile remnant contribution: dashed
lines; target remnant contribution: dotted lines;Pomeron contribution:
dashed dotted lines; sum: solid lines) and NA49 experiment \cite{Kadija:sqm2001}
(points). }}
\end{figure}

Fig.\ref{nexus3} depicts the rapidity spectra of baryons and antibaryons
from {\small NE}{\large X}{\small US} 3.0 with $P_{\mathrm{qq}}=0.02$
(solid lines). As a comparison, we also show the preliminary data
from the NA49 experiment \cite{Kadija:sqm2001} (points). The contributions
of projectile remnants (dashed lines), target remnants (dotted lines)
and cut Pomerons (dashed dotted lines) to particle production are
also shown respectively in Fig.\ref{nexus3}. Fig.\ref{nexus3} demonstrates
that {\small NE}{\large X}{\small US} 3.0 describes reasonably the
rapidity spectra of baryons and antibaryons in pp collision at 158
GeV. 

We also provide the particle yields at midrapidity, $y\in (y_{\mathrm{cm}}-0.5\, ,\, y_{\mathrm{cm}}+0.5)$,
from {\small NE}{\large X}{\small US} 3.0, Pythia 6.2 and compare
them to data in table \ref{table1}.%
\begin{table}[htp]
\begin{center}\begin{tabular}{|c|c|c|c|}
\hline 
yield&
{\small NE}{\large X}{\small US} 3.0&
Pythia 6.2&
NA49 data\\
\hline
\hline 
p&
9.12$\times $10$^{-2}$&
4.85$\times $10$^{-2}$&
9.28$\times $10$^{-2}$\\
\hline 
$\overline{\mathrm{p}}$&
2.00$\times $10$^{-2}$&
1.64$\times $10$^{-2}$&
2.05$\times $10$^{-2}$\\
\hline 
$\Lambda $&
1.61$\times $10$^{-2}$&
7.53$\times $10$^{-3}$&
1.79$\times $10$^{-2}$\\
\hline 
$\overline{\Lambda }$&
5.85$\times $10$^{-3}$&
4.02$\times $10$^{-3}$&
5.57$\times $10$^{-3}$\\
\hline 
$\Xi ^{-}$&
8.08$\times $10$^{-4}$&
2.53$\times $10$^{-4}$&
7.08$\times $10$^{-4}$\\
\hline 
$\overline{\Xi }^{+}$&
4.71$\times $10$^{-4}$&
2.20$\times $10$^{-4}$&
3.12$\times $10$^{-4}$\\
\hline 
$\Omega ^{-}$&
2.79$\times $10$^{-5}$ &
2.33$\times $10$^{-6}$ &
--\\
\hline 
$\overline{\Omega }^{+}$&
2.16$\times $10$^{-5}$ &
2.94$\times $10$^{-6}$ &
--\\
\hline
\end{tabular}\end{center}

\caption{\label{table1}Particle yields at midrapidity in pp collisions at
158 GeV. }
\end{table}

From {\small NE}{\large X}{\small US} 3.0, we get the ratios at midrapidity\[
\overline{\Xi }^{+}/\Xi ^{-}=0.58\, ,\, \, \, \, \, \, \, \, \, \, \, \, \, \, \, \, \, \overline{\Omega }^{+}/\Omega ^{-}=0.77\, .\]

In conclusion, it seems that old string models fail to reproduce the
experimental $\overline{\Xi }^{+}/\Xi ^{-}$ and anticipated $\overline{\Omega }^{+}/\Omega ^{-}$
ratio so far. The string formation mechanism as employed in {\small NE}{\large X}{\small US}
3.0 is able to reproduce the experimental data nicely. The rapidity
distributions of multi-strange baryons as well as $\Lambda $s and
protons can be understood. The main point is the fact that the final
result of a proton-proton scattering is a system of projectile and
target remnant and in addition (at least) one Pomeron represented
by two strings. At SPS energy, the soft Pomerons dominate. In general
the soft Pomerons are vacuum excitation and produce particles and
antiparticles equally because their string ends are sea quarks. The
valence quarks stay in remnants and favour the baryon production as
compared to antibaryon production.

\newpage
.

\end{document}